\documentclass[12pt]{wlscirep}

\usepackage[utf8]{inputenc}
\usepackage{eso-pic}
\usepackage{graphicx}
\usepackage{color}
\usepackage{amsmath}
\usepackage{caption}
\usepackage{setspace}
\usepackage{ulem}
\usepackage{hyperref}
\usepackage{lineno}
\usepackage[nomarkers,figuresonly]{endfloat}

\title{Predicting polymerization reactions via transfer learning using chemical language models}

\author[1]{Brenda S. Ferrari}
\author[2]{Matteo Manica}
\author[1,*]{Ronaldo Giro}
\author[2,3]{Teodoro Laino}
\author[1,*]{Mathias B. Steiner}
\affil[1]{IBM Research, Av. República do Chile, 330, CEP 20031-170, Rio de Janeiro, RJ, Brazil.}
\affil[2]{IBM Research Europe, Sa\"umerstrasse 4, 8803 R\"uschlikon, Switzerland.}
\affil[3]{National Center for Competence in Research-Catalysis (NCCR-Catalysis), Switzerland}

\begin{document}

\begin{abstract}

Polymers are candidate materials for a wide range of sustainability applications such as carbon capture and energy storage. However, computational polymer discovery lacks automated analysis of reaction pathways and stability assessment through retro-synthesis. Here, we report the first extension of transformer-based language models to polymerization reactions for both forward and retrosynthesis tasks. To that end, we have curated a  polymerization dataset for vinyl polymers covering reactions and retrosynthesis for representative homo-polymers and co-polymers. Overall, we obtain a forward model Top-4 accuracy of 80\% and a backward model Top-4 accuracy of 60\%. We further analyze the model performance with representative polymerization and retro-synthesis examples and evaluate its prediction quality from a materials science perspective.

\end{abstract}

\maketitle

\section*{Introduction}

Polymers have versatile properties and a wide range of applications~\cite{Arshad2020,Namazi2017,Patel2022}. The optimization of polymeric materials and the development of new polymers are, however, time-consuming processes. Machine Learning (ML) techniques have been demonstrated to significantly accelerate the discovery process by predicting polymer properties\cite{Kim2018,DoanTran2020} or, more recently, by enabling the automated design and generation of new polymers with predefined target properties\cite{Kim2021,Batra2020,Giro2023,park2023artificial}. Despite these advances, computational polymer discovery still faces major obstacles. Polymers are macromolecules which are formed by linking up smaller molecular units. Their synthesis typically involves various polymerization steps, with a multitude of possible links between monomer units. The prediction of thermodynamically stable polymer candidates, as well as the determination of a polymer's synthesizability\cite{Aziz2021}, is still affected by critical methodological limitations.

Recently, Caddeo et al.\cite{Caddeo2022} reported ML and atomistic approaches for modeling the thermodynamic stability of polymer blends while Chen et al.\cite{Chen2021} demonstrated a data-driven approach to automated retro-synthesis of target polymers. Kim et al.\cite{Kim2023} demonstrated the combination of ML-model based generation of new polymer candidates with a synthesizability analysis based on known polymerization reactions and commercially available reactants.

Despite the encouraging progress, significant gaps still exist in both methods and data domains. Currently, ML models do not exist for conducting retro-synthesis analysis on a range of co-polymers, polymer blends, ladder, cross-linked, and metal-containing polymers. Previous research has predominantly focused on homo-polymers, which can be easily represented as strings using the simplified molecular-input line-entry system (SMILES)\cite{Weininger1988,Weininger1989,Weininger1990}. The recent development of advanced string representations for polymers\cite{Lin2019,Lin2021} opens up new opportunities for modeling co-polymers\cite{Lin2019} as well as comb, branched, brushed, and star polymers\cite{Guo2022,Lin2021,Mohapatra2022,park2023artificial}.

Another critical issue is that the available polymer reaction datasets do not consider the influence of  solvents, catalysts, and experimental conditions. In addition, the data used to train ML models are not always made available publicly, compromising the reproducibility of model predictions. Overall, the lack of open data and open models severely hinders the advancement of computational polymer discovery.

\begin{figure}[!ht]
  \includegraphics[width=\linewidth]{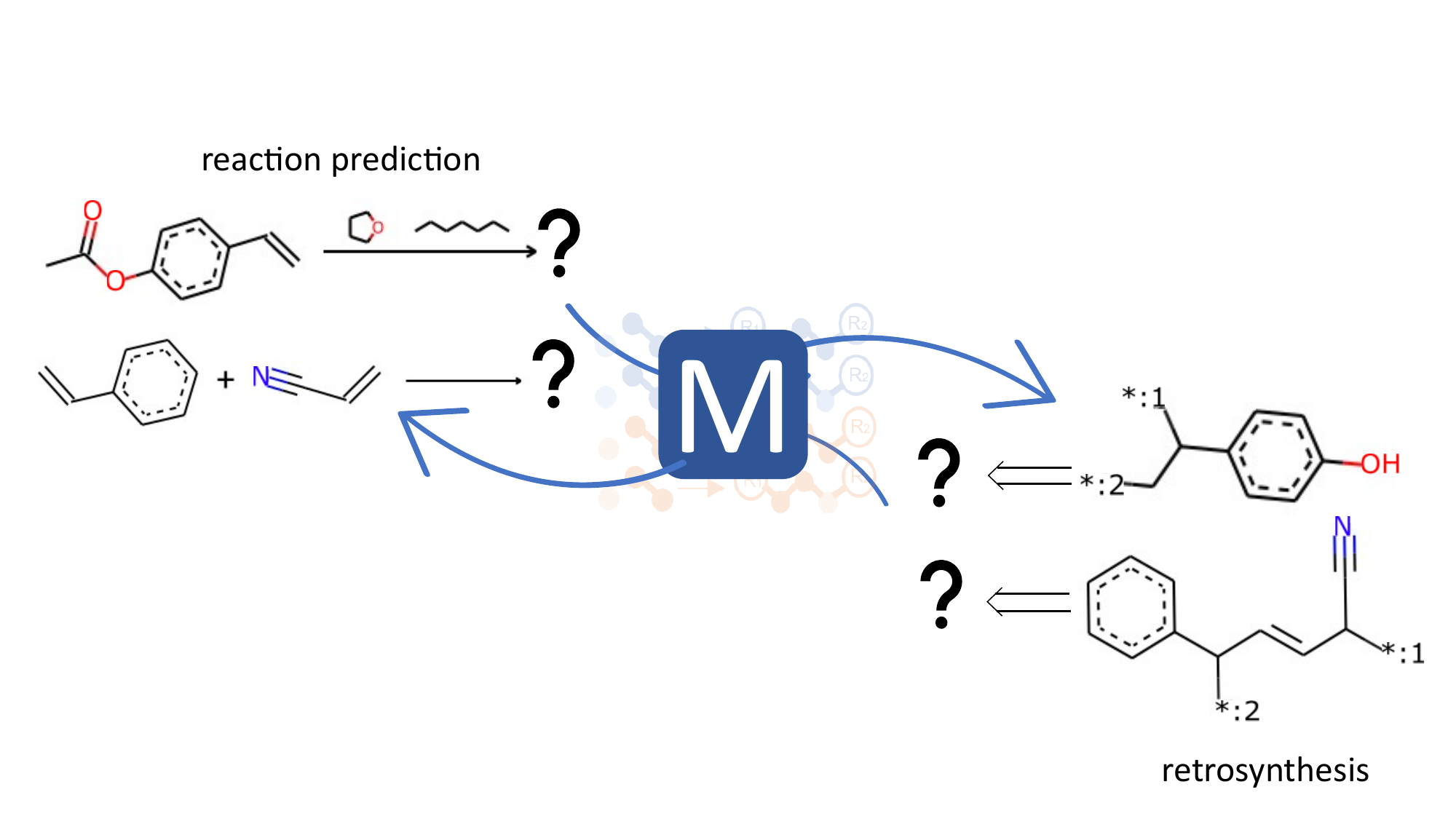}
  \caption{Problem representation. A Molecular Transformer model is being created for answering the following questions: "Given a set of reactants, which polymer could be obtained as product?" and "Given a certain polymer, how could it be synthesized?"}
  \label{fig1}
\end{figure}

In this work, we report the first extension of a transformer-based language model\cite{Schwaller2019,schwaller2020predicting} to polymerization reaction trained on a curated reaction dataset for vinyl polymers. We train the polymerization models for both forward and backward prediction tasks, addressing both homo-polymers and co-polymers consisting of up to two monomers. Our model predicts reactants, as well as reagents, solvents, and catalysts for each step of the retro-synthesis. Finally, we show that our models are able to perform two essential tasks as visualized in Fig.\ref{fig1}): (i) given a set of precursors, to predict a polymer product  and (ii) given a polymer, to suggest potential disconnections for synthetic strategies. To enable validation and reuse, we have made our models and data available in public repositories.

\section*{Results and Discussion}

\begin{figure}[!ht]
  \includegraphics[width=\linewidth]{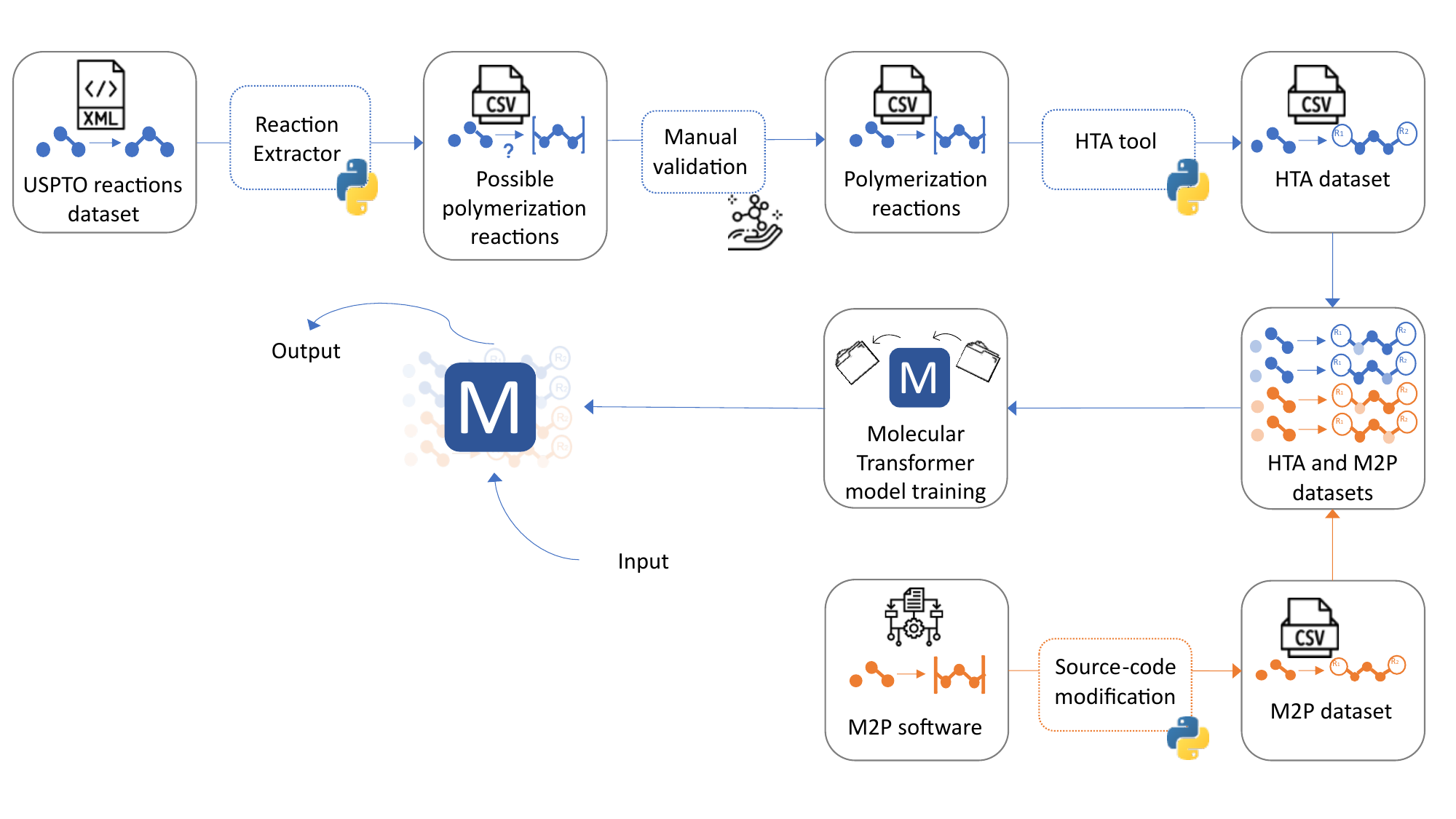}
  \caption{ Methodology flowchart. The workflow for predicting polymerization reactions (forward) and retro-synthesis analysis (backward) comprise data preparation and treatment, head and tail assignment with two different methodologies (HTA and M2P), model training and predictions in forward and backward directions.}
  \label{fig2}
\end{figure}

In Fig.\ref{fig2}, we visualize the end-to-end workflow for predicting polymerization reactions. The workflow includes dataset preparation and training of reaction and retrosynthesis prediction models, respectively. The training dataset was generated based on the publically available USPTO reaction dataset\cite{lowereactions,lowe2012extraction} which contains chemical reactions of organic compounds extracted from US patents issued between 1976 and 2016. For extracting polymerization reactions from the dataset, we have  designed a Python tool (see code availability section) that operates based on specific keywords. To ensure the selection of polymerization reactions only, we have employed a manual curation process that involves an individual review step of the reactions chosen by the automated procedure. Overall, we have analyzed 795 data entries for vinyl homo-polymers and co-polymers, respectively, resulting in two distinct datasets containing 3932 and 2965 reactions. These datasets cover all the possible combinations of the 795 reaction examples (details can be found in the  Methods section).

In general, polymer properties are determined to a large extent by how the monomer units are interconnected. For the purpose of our study, we have chosen linear chains as topological representations. For accurately predicting polymerization reactions, it is essential to correctly identify and label head and tail positions of the repeat units. To that end, we have adopted two distinct strategies. In the first approach, we have  adapted an existing tool for assigning head and tail atoms, referred to as Monomers-to-Polymer (M2P)\cite{wilson2020m2p}. In the second approach, we have developed a Python tool for Head-and-Tail assignment (HTA). We have provided extensive descriptions related to both HTA and M2P workflows in the Methods section. By using the two techniques, we have assigned head and tail atoms to constituent units within our polymer reaction dataset. We have then trained models on the two distinct datasets, labeled HTA and M2P, for comparative analysis of their predictive performance.

The modified M2P method can be applied to oligomers and assigns the positions of head and tail atoms in linkage bonds. The HTA method assigns head and tail atoms within monomers, thus defining the polymeric repeat unit. For facilitating the comparison of the ML models trained with the HTA and M2P datasets, respectively, we have also performed head and tail assignment in oligomers using the HTA routine. Throughout the training phase, the HTA dataset contained both monomers and oligomers, while the M2P dataset contained only oligomers. The inclusion of monomers within the HTA dataset enables the ML model to predict monomeric units of both homopolymers and copolymers. As the M2P dataset contains only oligomers, the respective model is not expected to predict homopolymer reactions correctly.

For reaction and retrosynthesis prediction modeling, we have used the Molecular Transformer architecture introduced by Schwaller et al.\cite{Schwaller2019,schwaller2020predicting}. In brief, the model is based on a vanilla transformer architecture\cite{vaswani2017attention} trained on textual representations of molecules. A Molecular Transformer casts chemical reaction prediction as a language modeling task\cite{Cadeddu2014}. We have encoded chemical reactions as sentences using reaction SMILES representation\cite{Weininger1988} of reactants, reagents as well as solvents and catalysts, along with the products. We have modeled forward- or retro-reaction predictions as a translation task from one language, i.e., reactants-reagents, to another language, i.e. products.
For training purposes, we have formally divided the reaction SMILES into source (reactants and reagents) and target (products) instances.
Since HTA and M2P datasets include different target outcomes for the same source instance, we have performed a splitting solely based on the targets. For model training, we have split the datasets on products in 95\% for training and 5\% for testing to ensure that no polymer (product) appears in both data sets.

To assess the performance of the Molecular Transformer trained on the two training datasets, we have used the Top-N accuracy metric for both forward and backward prediction models following the method reported in\cite{schwaller2020predicting}. We have calculated the model accuracy by considering the number of exact matches between the predicted canonical SMILES and the ground truth in the datasets. The Top-N accuracy considers that the ground truth canonical SMILES was found within the first N suggestions of the model. For example, if the ground truth target was found as the first suggestion in 70 out of 100 examples, it means Top-1 is 70\%. While round-trip is the generally preferred method for verifying the performance in the context of single-step retro-synthetic models\cite{schwaller2020predicting}, the datasets analyzed in our work link precursors to multiple products. In this case, the round-trip accuracy could be misleading, as multiple forward predictions are still valid for a precursor set and multiple products map to the same precursors. To avoid this, we have used Top-N accuracy for evaluating the performance of both forward and backward models.

\begin{figure}[!ht]
  \includegraphics[width=\linewidth]{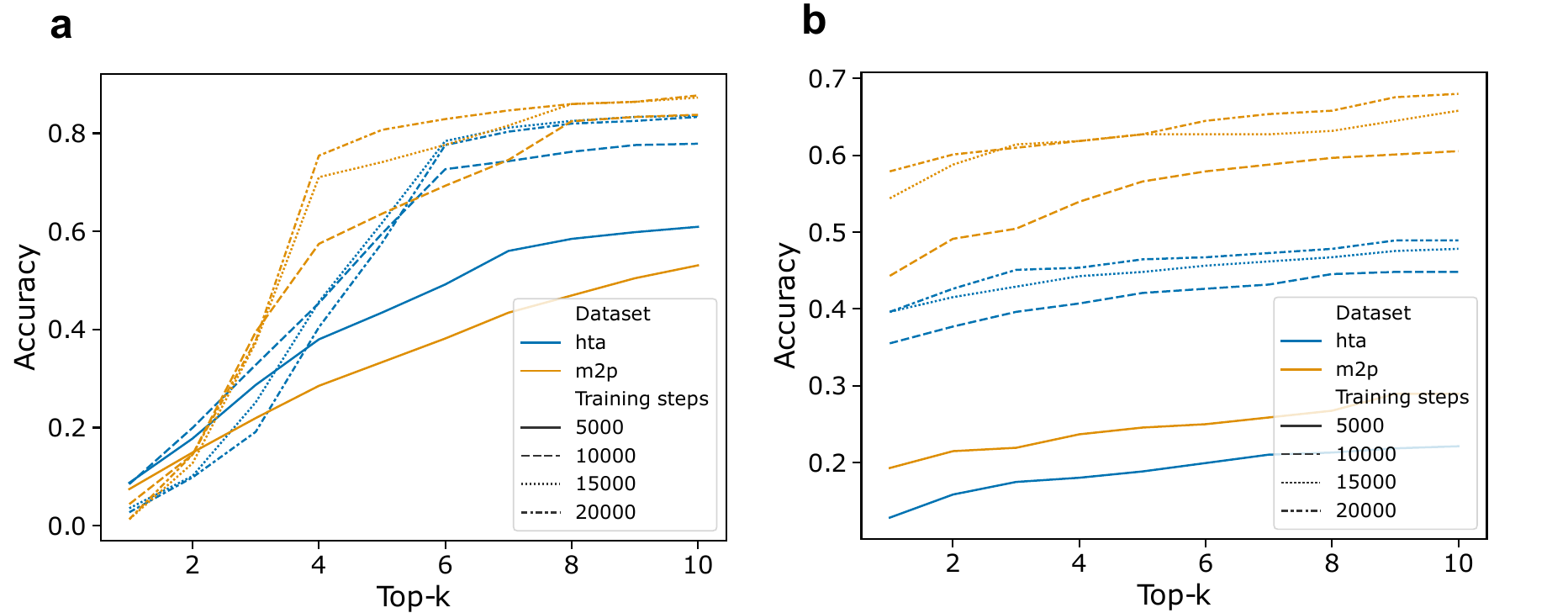}
  \caption{Prediction model performance. a) Polymerization reaction prediction (forward model) accuracy. b) Retro-synthesis prediction (backward model) accuracy.}
  \label{fig3}
\end{figure}

In Fig.\ref{fig3}, we show the prediction model performance obtained for the two datasets. The M2P dataset shows better performance overall in both forward and backward models, see Fig.\ref{fig3}a-b. In backward predictions, we observe the general trend that the higher the number of training steps, the higher the model accuracy. For forward predictions, this trend only manifests in certain intervals of the Top-N range. The accuracy increases monotonously in both forward and backward modes, albeit with different slopes. We observe a sharp accuracy increase in forward model for M2P around Top-3 and HTA around Top-4, respectively. This could be explained by the number of possible reaction outcomes. While M2P provides $n$ reaction outcomes as oligomers built from combination of reagent monomers, HTA also provides the repeat units as product of polymerization. This means that HTA provides $n+1$ or $n+2$ results, depending on the number of reagent monomers involved in the reaction. On average, M2P returns 4 possible reaction outcomes while HTA returns 5 or 6.

The observation that the M2P dataset yields superior model performance could be due to the simpler learning process of polymerization rules within this dataset. The M2P algorithm polymerizes monomers in all possible functional groups and chooses a representative structure randomly. Due to the random character of the M2P algorithm, different realizations result in different choices of representative structures, affecting the ML training performance. In comparison, the HTA algorithm identifies reactive sites through the analysis of nucleophile and electrophile atoms, applying the Mulliken’s scheme\cite{mulliken1955electronicI,mulliken1955electronicII,mulliken1955electronicIV} for identifying the most probable structure relating to chemical rules. In other words, M2P structures are a combination of all possible bond connections between monomers, while HTA structures are combinations of all possible connections between reacting sites.

\begin{figure}[!ht]
  \includegraphics[width=15cm]{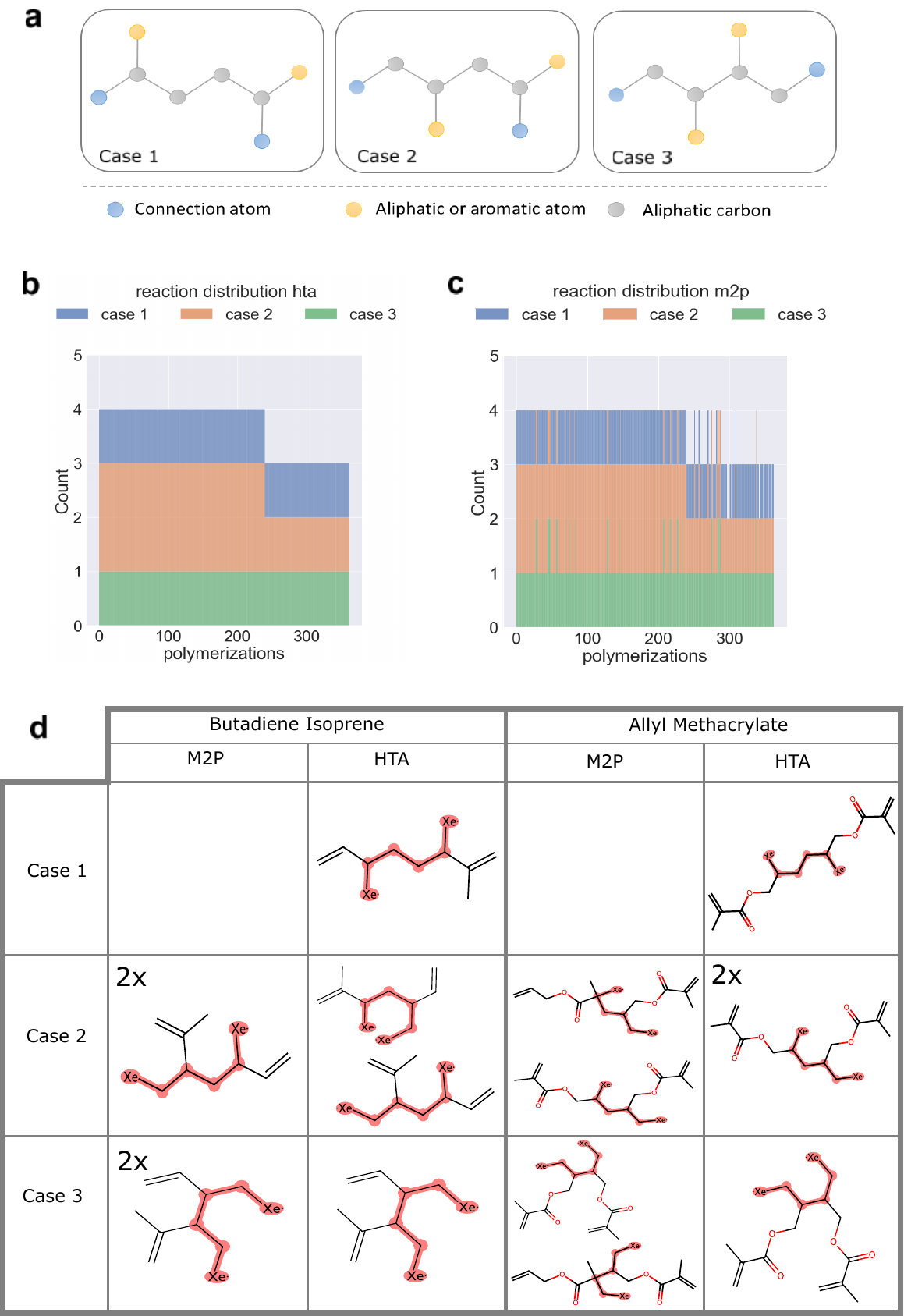}
  \caption{Data representation. "2X" representation means the same structure appears twice.  a) SMARTS representation of the vinyl bond formation. b) Comparative distribution of HTA data. c) Comparative distribution of M2P data. d) Examples of Butadiene Isoprene and Allyl Methacrylate.}
  \label{fig4}
\end{figure}

To clarify this point, let us consider how the repeat units in the HTA dataset are linked up to form oligomers. A bond between two vinyl monomers with only secondary carbon atoms may be formed as visualized in the example shown in Fig.\ref{fig4}a. We note that the polymeric repeat unit generated by HTA was considered for inclusion into the dataset, however, it was disregarded in the distribution analysis. This is also the case for oligomers with tertiary carbons.

In case 1, the bond is formed between the carbon atoms at the end of the monomers in the chain. As a result, both head and tail are localized at external atoms of the reaction site. We refer to this connection type as tail-tail. In case 2, head and tail are localized at internal and external carbon positions, respectively. We refer to this connection type as head-tail. Finally, in case 3, the bond occurs between secondary carbon atoms of the double bond. Once polymerized, both head and tail atoms are located at internal carbon atom sites. We refer to this connection type as head-head. By analyzing the case distribution in the dataset for model training, see Fig.\ref{fig4}b, we find that the HTA dataset contains 1/3 of each case for oligomers with 3 different combinations while the ratio is 1/2/1 for oligomers with 4 different combinations. The latter can be explained by the twofold possibility in case 2 of bond formation due to the presence of two monomers. Note, that the M2P dataset does not have a fixed case ratio. This is because M2P performs the polymerization for all possible functional groups of the molecular structure, see Fig.\ref{fig4}c.

Those differences on the distribution are observed on examples in Fig.\ref{fig4}d. For the butadiene isoprene polymer with its four potential polymerizations, the vinyl bond case ratio 1/2/3 representing cases 1, 2 and 3, respectively, see Fig.\ref{fig4}a, is 1/2/1 for HTA and 0/2/2 for M2P. Similarly, in the case of allyl methacrylate, we obtain the case ratio 1/2/1 for HTA and 0/2/2 for M2P. In case of M2P, the polymerization is performed by considering all the functional groups of the monomer. The results observed in Fig.\ref{fig3}a-b could indicate that the model has learned this pattern efficiently. The larger spread of accuracy values observed in the retro-synthesis model could be due to the specifics of the oligomers.

While we obtain overall better modeling results with M2P, both datasets reveal interesting insights. Despite showing a Top-1 accuracy below 10\%, the forward model exhibits Top-4 and Top-6 accuracy around 80\%, which suggests a direct relation with the way the two datasets have been compiled. Indeed, by construction, the same set of reactants are associated with multiple polymers.
The backward model has a Top-1 accuracy of about 60\% for M2P and 40\% for HTA. The lower accuracy observed in HTA could be explained by the ease that the model may have learned the polymerization pattern represented in M2P data, as explained previously. We will expand this analysis in the following paragraphs by investigating the usefulness of the model outputs from a materials science perspective. 

For our domain applicability analysis, see Methods section for details, we have selected representative polymers from the literature \cite{Saleh2007,FranciscoVieira2019,Concilio2021, Atta2013,Chen2000,Ogieglo2015,dena2020surface,Ibrahim2003}. A comparison of these reactions reveal product similarities ranging from 0 to 30\% for HTA and M2P datasets while reactants similarities range from 0 to 12\%, see Supplemental Table S1. Co-polymers show increased similarity values in M2P, about 3-6\% higher, attesting to their representation in the training data. Homo-polymers exhibit increased similarity of about 4\% in HTA as the dataset includes monomer representations.

Overall, both models correctly predicted  6 out of 8 reactions in Top-4 and could suggest at least one correct monomer in all the examples studied. The HTA based model correctly predicted 3 out of 8 reactions in Top-1 and 4 out of 8 reactions in Top-4, while the M2P based model correctly predicted 1 out of 8 reactions in Top-1 and 2 out of 8 reactions in Top-4. Note, that the HTA based model predominantly matches homo-polymers while M2P matches mainly co-polymers. The pattern is plausible as HTA contains the monomers of all polymers while M2P does only contain oligomers.

\begin{figure}[!ht]
  \includegraphics[width=\linewidth]{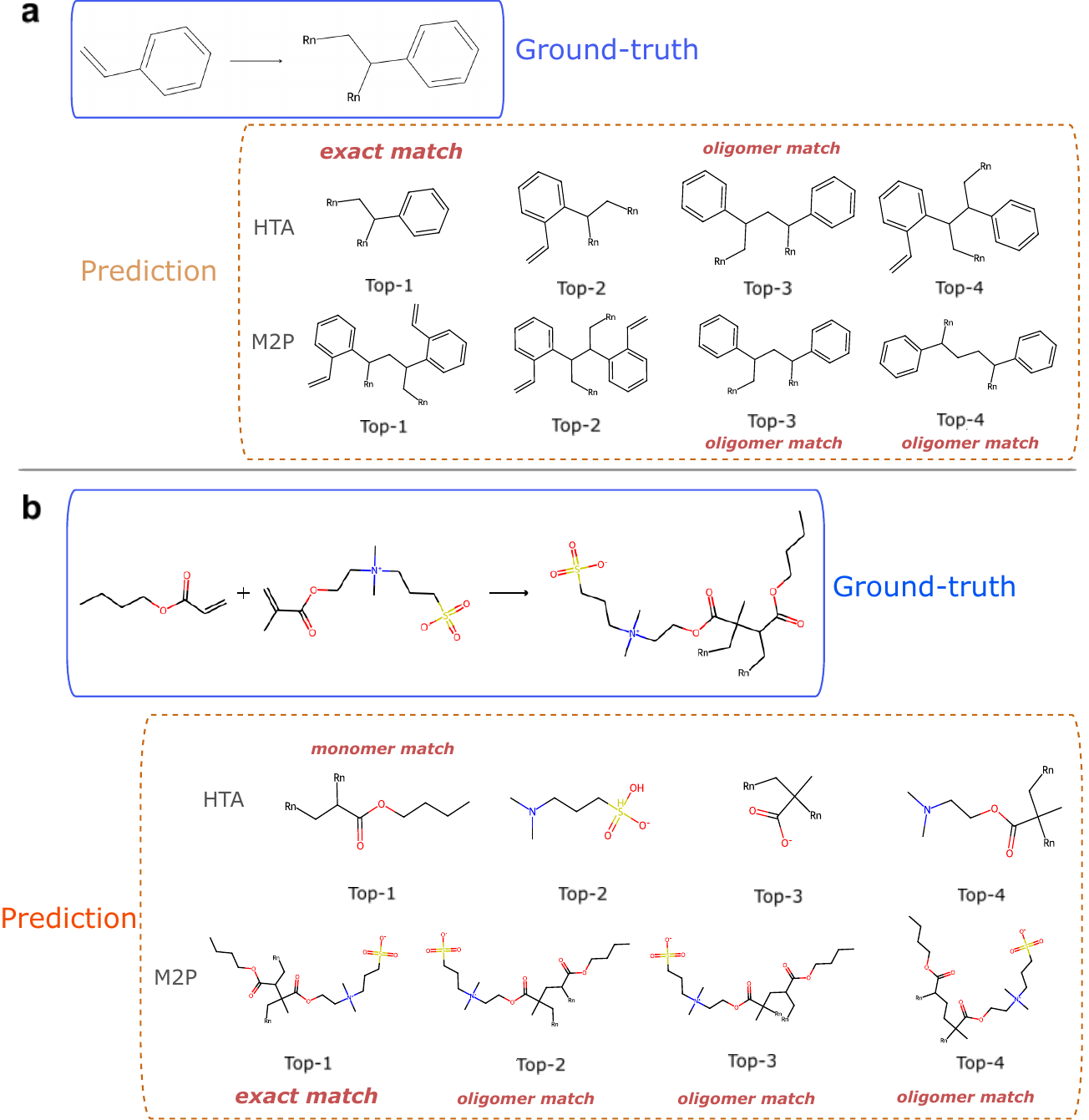}
  \caption{Representative examples. Model predictions using the Molecular Transformer trained on HTA and M2P datasets, respectively. Catalysts, solvents, and stochiometry are not shown. a) Polystyrene. b) p(SBMA-nBA) copolymer. In 2D molecules representations carbon atoms are in black, oxygen and hydroxyl in red, nitrogen in dark blue, and sulfur in yellow. The connection points of polymer repeat units are represented with Rn atoms.}
  \label{fig5}
\end{figure}

For the polymerization example of styrene, see figure \ref{fig5}a), the HTA based model achieves a full SMILES match at Top-1 as well as the representation of a possible oligomer structure, with 2 connected repeat units, at Top-3. In case of the M2P based model, we do not obtain a match for the actual product. The oligomer representation is shown for Top-3 and Top-4. For the polymerization of the co-polymer p(SBMA-nBA), see Figure\ref{fig5}b, the model predicts an exact product match for Top-1, along with the all other bond formation possibilities on Top-2 to Top-4. This means that the model is able to correctly predict the connections in the polymerization reactions. While the HTA model failed to predict the actual result, the model was able to identify the correct head and tail positions of one of the repeat units (Top-1). In addition, the model suggested fragments of the monomer seen as Top-2 and Top-4.

One interesting exception is shown in Supplemental Fig.S1b. In the polymerization of p(xMA), a co-polymer, both models suggested incorrect structures at Top-1. However, the HTA based model generates the correct repeat units for all four predictions, Top-2 being the exact match. The M2P based model merely predicts all possible links between carbon atoms for generating the polymeric bond, and one of the monomers is an exact match. For p(St-BuA), see Supplemental Material, Fig.S1b, the HTA based model predicts the correct repeat units in Top-1 and Top-2. As expected, however, it fails to generate the oligomer. Nevertheless, the M2P based model predicts the correct monomers and the exact match is shown in Top-4. 

In the example of Polyvinyl chloride polymerization, see Supplemental Material, Fig.S2a, we observe an interesting model behavior. While neither HTA nor M2P data underwent special processing for monomers/oligomers with protection groups, the model learned to predict output without the protection group. The HTA based model suggested the correct structure for polyvinyl chloride at Top-1, without the protection group. The M2P based model, however, failed to generated an output that resembled the ground-truth structure. In the polymerization of p(DOM-DVB), see Supplemental Fig.S2b, we observe that both models struggles to predict polymers in which monomers have the double bond in the middle of the chain. Nevertheless, both models correctly suggested one of the monomers and its bonds combinations.

Both models correctly predict oligomers formed by monomeric units with halogens, such as chlorine. Since all training data is tagged with a token (Rn) representing the location for the continuation of the chain, all model predictions suggest the formation of monomers with that token in its structure. This is shown in Supplemental Figure S3a for the polymerization of p(tC-tBuM) copolymer. The HTA based model accurately predicts one of the monomers and its combinations while the M2P based model fails this task. Even in the presence of a large number of reactants, catalysts, and solvents, the model is able to correctly predict the polymers, as shown in case of Poly(n-butyl methacrylate), see Supplemental Fig.S3b. As expected for homo-polymers, the HTA model predicts the exact match in Top-1 along with some monomer combinations in Top-2 and Top-3 while the M2P based model predicts the combinations of the monomer in Top-1 to Top-3.

For the curated examples, the HTA based model predicts a higher number of exact matches for the polymer structures in Top-1 (3 out of 8) and Top-4 (4 out of 8), respectively. In cases of incorrect predictions, the model delivered at least one of the monomers correctly. The model trained with M2P data had limitations regarding homo-polymers, as expected. Nevertheless, the M2P model correctly predicts complex co-polymers and a very close match for p(tC-tBuM) copolymer, a pattern not represented in the training dataset. Both models appear to have complementary performance, predicting exact matches for 6 out of 8 reactions and suggesting at least one correct monomer for all the examples studied. For increasing the likelihood of a suitable prediction outcome, we, therefore, recommend the joint utilization of both HTA and M2P based models for domain specific applications 

\section*{Conclusion}

In summary, we have reported the curation of a vinyl polymerization reaction dataset and the training of a Molecular Transformer algorithm for predicting polymerization (forward) and retro-synthesis (backward) reactions. For dataset curation, we have introduced two novel algorithms for assigning head and tail positions, named HTA and M2P. We have applied both algorithms to process 795 data entries for vinyl homo-polymers and co-polymers and produced two separate datasets with 3932 and 2965 reactions, respectively, representing all possible combinations of the 795 reaction examples. Upon training, the Molecular Transformer exhibits a forward-model (Top-4 and Top-6) accuracy around 80\% for both datasets. The retro-model exhibits a Top-1 accuracy of about 60\% for the M2P dataset and 40\% for the HTA dataset. 

We have showcased the capabilities of the models through a case study involving eight reactions. These reactions were selected based on examples provided in the literature. Both models have predicted 6 out of 8 reactions as exact match at Top-4, and suggested at least one correct monomer for all the examples studied. The models work in a complementary manner, as the model trained with the HTA dataset produces better results for homo-polymers while the model trained with the M2P dataset predicts better matches for co-polymers.

 Based on our analysis  of the strengths and limitations of the Molecular Transformer approach, we expect that extending the model training to include other polymer classes will broaden  model applicability and further increase the robustness of prediction outcomes. The lack of available data on polymerization reactions and tools for head and tail assignment were major challenges we have encountered in this work. Therefore, we have made our curated datasets and tools  publicly available for reuse and validation. 

\section*{Methods}

\subsection*{Polymerization dataset}

The polymerization reactions and polymer names were extracted from a publicly available dataset \cite{lowereactions} derived from the patent mining work of Lowe\cite{lowe2012extraction}. This dataset is composed by approximately 1.8M chemical reactions, extracted from 1976 to September 2016 USPTO granted patents. A Python script was developed to automate the data extraction. Only chemical reactions and molecule names that presented the keyword “polymerization” on the experimental procedure text were chosen. After the automated step, a manual validation was performed to remove data entries in which the “polymerization” keyword was related to any information not compatible with the reaction type. In this step the number of data points were reduced from 8.668 to 3.286 possible polymerization reactions. In the Lowe\cite{lowe2012extraction} dataset, the head and tail atom positions to define the polymer repeat units of polymerization reactions products are missing. How these monomers are linked play an important role in polymer properties\cite{zhou2011preparation}. Since there was no established methodology to perform the assignment of the head and tail in polymer structures represented by SMILES notation, Python tools with two different approaches were developed to perform this task. In the first approach we used an in house developed Python tool, called HTA (Head-and-Tail Assignment), to assign the head and tail atoms (more details see Methods section). In the second approach a modified version of Monomers-to-Polymer (M2P)\cite{wilson2020m2p} tool was developed to assign the head and tail atoms. These two approaches resulted in two datasets, composed by 795 data entries, related to vinyl homo-polymers and co-polymers with 2 monomer and were properly clean from duplicates and erroneous reactions. Besides the head and tail assignment, another two datasets were generated by describing all the possible product outcomes which are represented by one or two products and the different bond formation between the monomers. The bond formations were performed by the combination of monomers using rdkit.Chem.rdChemReactions method. For that, all the monomers combination were considered according to M2P and HTA algorithms. On the HTA algorithm the monomers were also considered as possible outcome of the reaction. In this sense, regarding the number of results m2p=n and hta=n+1/n+2. This increased the number of reactions from 795 to 3932 and 2965 reactions, for HTA and M2P respectively. In summary, four datasets were generated and two datasets were used to train our model: the all monomers combination datasets for HTA and M2P.

\subsection*{Data distribution}

Both M2P and HTA datasets were sorted by polymer name and repeating unit, the latter alphabetically and by length. All the results for the same polymer were grouped in lists during pre-processing process. The modified M2P tool assign the head and tail atom positions (linkage bounds) in oligomers, while the HTA tool in the monomer, defining the polymeric repeat unit. With the purpose to avoid any bias during the ML training model between the two datasets, we also considered the head and tail assignment with the HTA tool in oligomers. This fact adds another level of complexity: how the repeat units are linked. There are three possible cases: (i) tail-tail; (ii) head-tail and (iii) head-head. For the extraction of the distribution of cases, there were set SMARTS\cite{smarts} for each polymerization case and after a dearomatization process, all the SMILES\cite{Weininger1988} were compared to the SMARTS set, using the RDKit\cite{rdkit} library. SMARTS\cite{smarts} is a chemical structure query language for describing molecule patterns. RDKit can import SMARTS queries for use in searching of SMILES patterns. Cases that deviated from the standard SMARTS query pattern (i.e., tertiary carbons that could cause uncertainties on the algorithm) were not considered. After post-processing, both datasets were merged, since only equal polymers were considered on the comparison, and a distribution chart was built with the results.

\subsection*{Applicability domain analysis}

The polymers that were used on this case study were manually extracted from the literature \cite{Saleh2007,FranciscoVieira2019,Concilio2021, Atta2013,Chen2000,Ogieglo2015,dena2020surface,Ibrahim2003}. The SMILES representation of polymers were canonicalized using the RDKit\cite{rdkit} package. The fingerprint calculation was performed by defining the fingerprints of the input data and the data used on the Molecular Transformer training using RDKFingerprint\cite{rdkit} followed by the comparison between both datasets. Each input data fingerprint was compared with the fingerprints of the whole training data. The results obtained comprised on the mean of the comparison results and the maximum value on the list. This process was performed separately for reactants/reagents and products.

\subsection*{HTA algorithm}

For the head and tail assignment using the HeadTailAssigner (HTA) tool, the reaction SMILES was used as input. However, the algorithm also accepts monomer SMILES as input. Following the pre-processing analysis, the most probable monomer in the reaction string was defined by comparing the products with the reactants. The last step was performed by a fingerprint similarity analysis, using the RDKFingerprint\cite{rdkit} and maxPath=7 and a comparison using Tanimoto Similarity \cite{rdkit, tanimoto1958elementary}. The vinyl class is the focus of this work, but the algorithm may also identify and assign head and tail of polyamides, polyesters, polyurethanes and polyethers. To define the polymer class, the algorithm searches all the possible functional groups on the molecular structure by substructure match with the SMARTS pattern of each organic function. In a next step, it compares the atomic index of nucleophilicity\cite{Szczepanik2013} and the functional groups extracted from the monomer. If the monomer smiles has only one functional group, a SMARTS pattern is acquired to classify the polymerization mechanism. If the monomer smiles has two or more functional groups, the priority of polymerization is decided based on the atomic index of nucleophilicity\cite{Szczepanik2013}. The atomic index of nucleophilicity of an atom X involving only the highest occupied molecular orbital (HOMO) $n$ is defined as\cite{Szczepanik2013}:

\begin{equation}
    R_X = \frac{\sum _\alpha^X\left |C_{\alpha,n}  \right |^2}{(1 - \varepsilon_n)}
\end{equation}

\noindent
where $C_{\alpha,n}$ are the molecular orbital expansion coefficients of $\alpha$th atomic orbital on molecular orbital $n$ (HOMO) and $\varepsilon_n$ is the HOMO energy.

The $R_X$ was calculated within STO-3G basis set and with the Mulliken's population analysis\cite{mulliken1955electronicI,mulliken1955electronicII,mulliken1955electronicIV} scheme. All the quantum states functions were calculated at RHF theory level, using the standard \textit{ab initio} quantum chemistry package GAMESS\cite{Barca2020} version 2021 R2.

In summary, the higher the atomic population value in an atom, higher the atom index of nucleophilicity $R_X$, which means, the atom has more probability on being the polymerization site\cite{Szczepanik2013}. The condition is set depending on the relation between polymerization class and the functional groups present in the structure. If one atom has a higher $R_X$ but its functional group is not represented in any polymer class, the algorithm is going to keep searching until it finds an atom that is represented in an existing polymer class. After obtaining a match, the functional groups are concatenated up until it is a match with a previously defined class. The mechanism is defined depending on the polymer class described previously. If the class is vinyl and the algorithm detects the presence of an specific catalyst, it may also define if the mechanism is anionic, cationic or radicalar. With all the information obtained previously, the algorithm defines the head and tail by assigning the atom id of the respective nucleophile and electrophile on the functional group responsible for the polymerization.

\subsection*{M2P algorithm}

For the head and tail assignment using Monomers to Polymers (M2P)\cite{m2p2022github}, a modified version of the M2P algorithm was used. According to the authors "The library can generate multiple replicate structures to create polymer chains represented at the atom and bond level. RDKit\cite{rdkit} reaction SMARTS\cite{smarts} are used to manipulate the molecular structures and perform in silico reactions. The polymer chemistries available include vinyls, acrylates, esters, amides, imides, and carbonates."\cite{m2p2022github}. From the source-code, the algorithm was modified to generate head and tail assignments for vinyl polymerization only if the user checks True for the head and tail creation parameter. The vinyl polymerization comprises the initiation, propagation and termination steps with token atoms (Kr, Xe and Rn) used on the reaction SMARTS to define the bond formation site. In the end of the polymerization, these tokens would be deleted, to keep only the polymer product as a result. For the modified version, the token atoms were added on the initiation, propagation and termination step to represent the formation of the head and tail atoms on the polymer. In the end of the polymerization process, these tokens remain on the polymers to represent the head and tail assignment. This treatment was also extended for co-polymers with 3 monomers.

\subsection*{Model training for forward and backward reaction prediction}

As model, for both forward and backward reaction prediction, we considered the Molecular Transformer proposed by Schwaller et al.~{\cite{Schwaller2019}}.
Encoders follow a standard \textit{transformer} architecture with $6$ layers, word vectors and RNN decoders of size $512$, the gradient was accumulated 8 times with a maximum vector norm of $0.0$, and \textit{adam} was used as an optimizer ($\beta_{1}=0.9$, $\beta_{2}=0.998$). Batch size was set to $4096$, and the batch type as well as the gradient normalisation method to \textit{tokens}. The learning rate was set to 2.0 with \textit{noam} as decay method. Dropout and label smoothing ($\epsilon$) were set to $0.1$. Parameter initialisation was disabled and position encoding enabled. All models were trained using a version of OpenNMT~{\cite{klein-etal-2017-opennmt}} adapted for the Molecular Transformer~{\cite{rxnonmt}}.
Compared to the standard Molecular Transformers we extended the model and tokenizer to handle head and tail representations using noble gasses as additonal tokens. We trained models on the datasets generated both with the HTA and the M2P algorithm ad compared the both backward and forward performance.

\section*{Data Availability}

D. Lowe's dataset 1976\textunderscore Sep2016\textunderscore USPTOgrants\textunderscore cml.7z used to extract the polymerization reaction data is available under the doi:10.6084/m9.figshare.5104873.v1 - at

\noindent
\url{https://figshare.com/articles/dataset/Chemical_reactions_from_US_patents_1976-Sep2016_/5104873?file=8664364}

The training dataset file hta\textunderscore dataset\textunderscore all\textunderscore combinations.csv containing polymerization reactions in SMILES format and with the head and tail atoms assigned by the Python tool HTA is available under the doi:10.24435/materialscloud:zw-be - at

\noindent
\url{https://archive.materialscloud.org/record/2023.137}

The training dataset file m2p\textunderscore dataset\textunderscore all\textunderscore combinations.csv containing polymerization reactions in SMILES format and with the head and tail atoms assigned by the modified version of M2P Python tool is available under the doi:10.24435/materialscloud:zw-be - at

\noindent
\url{https://archive.materialscloud.org/record/2023.137}

The file trained\textunderscore models.zip contains the Machine Learning training models (forward and retrosynthesis) and is available under the 
doi:10.24435/materialscloud:zw-be - at

\noindent
\url{https://archive.materialscloud.org/record/2023.137}

\section*{Code Availability}
The code for extracting the polymerization reaction data from Daniel Lowe's dataset is available at: \url{https://github.com/IBM/XLMExtractor-chem-reaction}.

The code for assigning the head and tail atoms using quantum chemistry and polymerization mechanisms information is available at:\url{https://github.com/IBM/HeadTailAssign}.

The code for assigning the head and tail atoms based on the Monomers to Polymers (M2P) tool is available at: \url{https://github.com/IBM/m2o-head-tail-assign}

The code for model training is available at: \url{https://github.com/rxn4chemistry}.


\section*{Acknowledgements}
T. L. acknowledges support from the NCCR Catalysis (grant number 180544), a National Centre of Competence in Research funded by the Swiss National Science Foundation. 

\section*{Author Contributions}
 B. S. F. created and curated the polymerization reaction dataset and  co-wrote the manuscript. 
 M. M developed Machine-Learning models and co-wrote the manuscript.
 R. G. conceived the work and co-wrote the manuscript.
 T. L. conceived the work and co-wrote the manuscript.
 M. B. S. conceived the work and co-wrote the manuscript.

\section*{Competing financial interests:} The authors declare no competing financial interests.

\section*{Additional Information}
\subsection*{Supplementary information}
Supplementary Information, including Supplementary Table S1 and Supplementary Figures S1-S4, are available as a pdf-file

\subsection*{Correspondence} and requests for materials should be addressed to mathiast@br.ibm.com

\end{document}


\section*{SUPPLEMENTARY INFORMATION}

\subsection*{Results and Discussion}

We provide supplemental table and figures to support the discussion on the main manuscript. These figures demonstrate the performance of the trained Molecular Transformer mode. The polymers that were considered on these case studies were manually extracted
from the literature followed by the applicability domain analysis (for more details see Methods section).

\bibliography{main}

\newpage

\begin{figure}[!ht]
  \includegraphics[width=\linewidth]{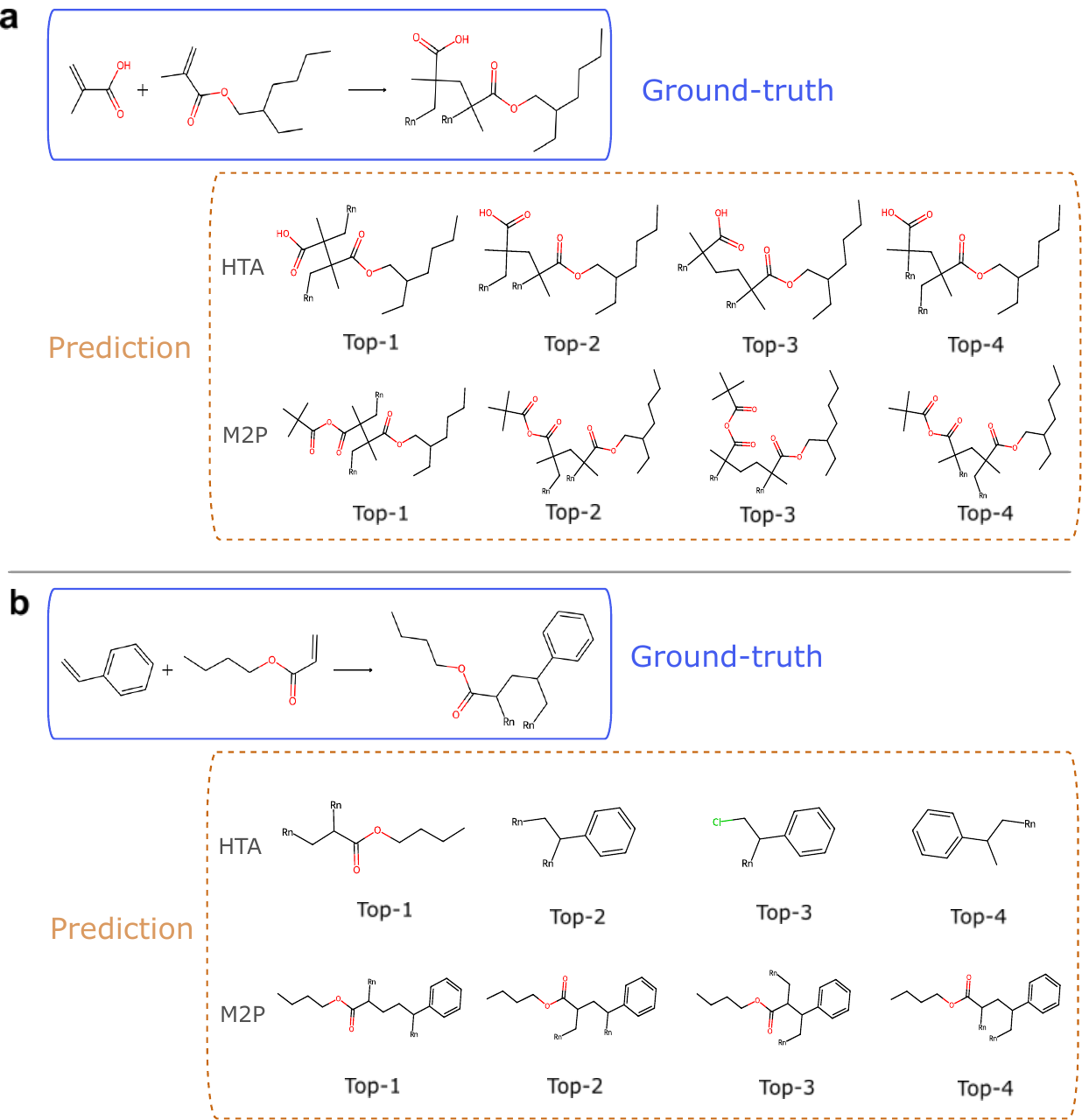}
  \caption{ Example of prediction using Molecular Transformer model trained with HTA and M2P data. Catalysts, solvents and stoichiometry not shown. a) p(xMA) copolymer. b) p(St-BuA) copolymer. In 2D molecules representations carbon atoms are in black and oxygen in red. The connection points of polymer repeat units are represented with Rn atoms.}
  \label{fig6}
\end{figure}

\begin{figure}[!ht]
  \includegraphics[width=\linewidth]{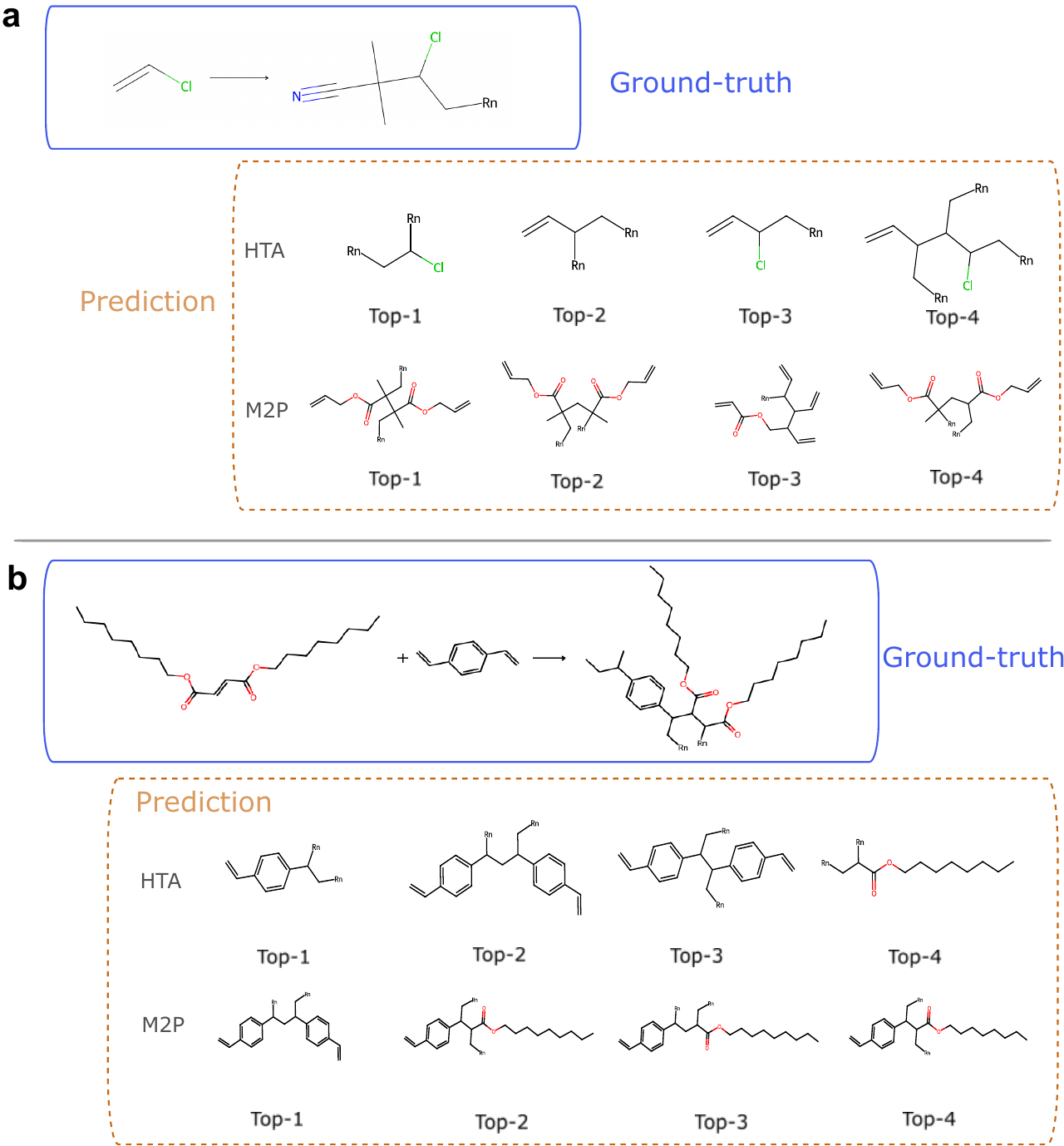}
  \caption{ Example of prediction using Molecular Transformer model trained with HTA and M2P data. Catalysts, solvents and stoichiometry not shown. a) Polyvinyl chloride. b) p(DOM-DVB) copolymer. In 2D molecules representations carbon atoms are in black, oxygen in red, nitrogen in dark blue, and chloride in green. The connection points of polymer repeat units are represented with Rn atoms.}
  \label{fig7}
\end{figure}

\begin{figure}[!ht]
  \includegraphics[width=\linewidth]{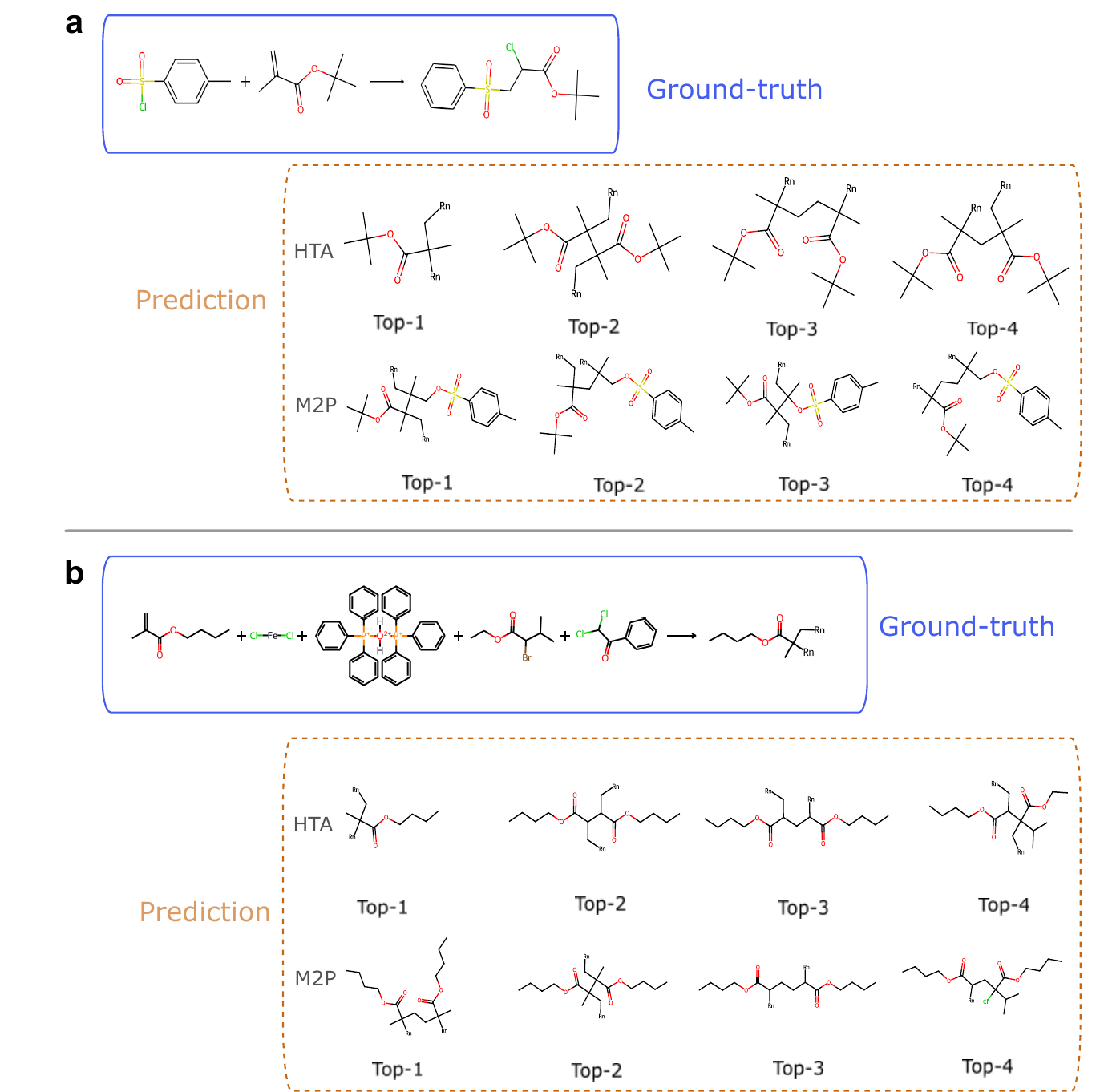}
  \caption{Example of prediction using Molecular Transformer model trained with HTA and M2P data. Catalysts, solvents and stoichiometry not shown. a) p(tC-tBuM) copolymer. b) Poly(n-butyl methacrylate). In 2D molecules representations carbon atoms are in black, oxygen in red, chloride in green, phosphorus in orange, boron in brown and sulfur in yellow. The connection points of polymer repeat units are represented with Rn atoms.}
  \label{fig8}
\end{figure}

\begin{table}
  \includegraphics[width=\linewidth]{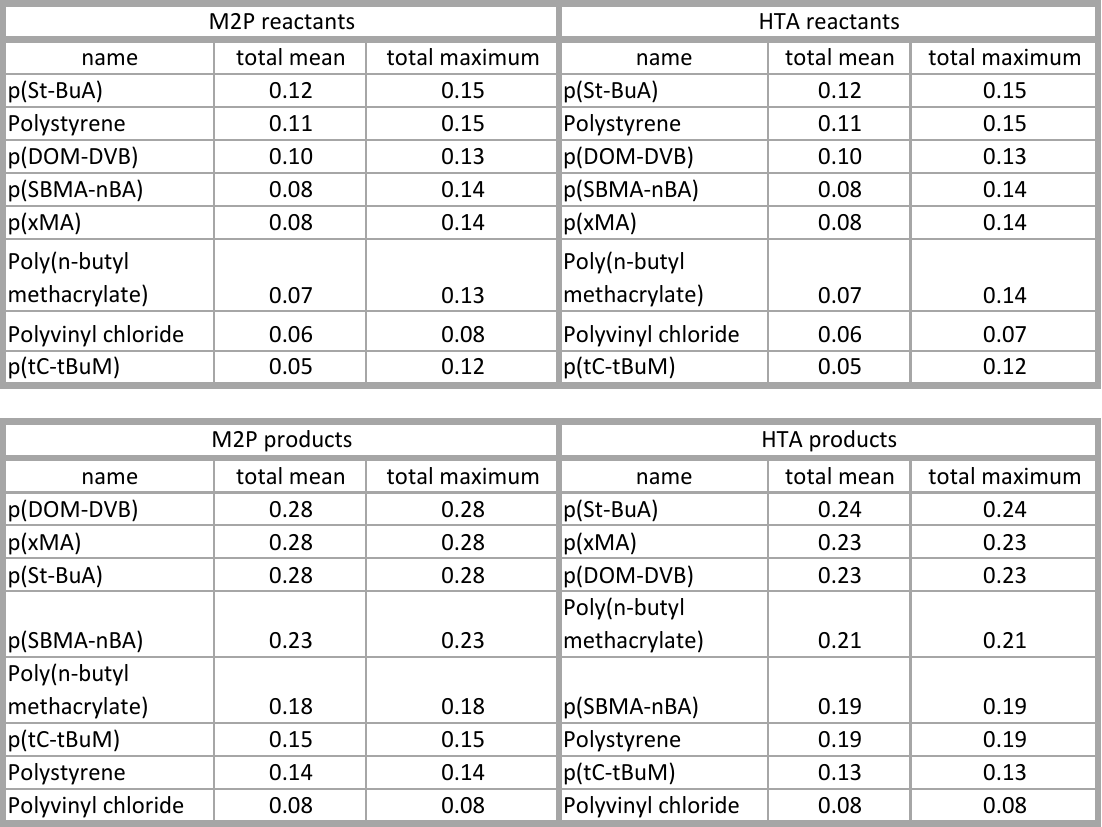}
  \caption{Applicability domain analysis results for HTA and M2P datasets.}
  \label{suptable1}
\end{table}